\documentclass[12pt,preprint]{aastex}
\renewcommand{\cite}{\citealp}

\newcommand{\msol}{$M_\odot$}

\shorttitle{The Leo\,IV dSph galaxy}
\shortauthors{Moretti et al.}

\begin{document}


\title{The Leo\,IV dwarf spheroidal galaxy: color$-$magnitude diagram and pulsating stars\altaffilmark{1}}


\author{Maria Ida Moretti\altaffilmark{2},
Massimo Dall'Ora\altaffilmark{3},
Vincenzo Ripepi\altaffilmark{3},
Gisella Clementini\altaffilmark{4},
Luca Di Fabrizio\altaffilmark{5},
Horace A. Smith\altaffilmark{6},
Nathan De Lee\altaffilmark{7},
Charles Kuehn\altaffilmark{6},
M\'arcio Catelan\altaffilmark{8,9,10}, 
Marcella Marconi\altaffilmark{3},
Ilaria Musella\altaffilmark{3},
Timothy C. Beers\altaffilmark{6,11},
Karen Kinemuchi\altaffilmark{7,12}
}

\altaffiltext{1}{Based on data collected at the 2.5 m Isaac Newton Telescope, La Palma, Canary Island, Spain, 
at the 4.2 m William Herschel Telescope, Roche de los Muchachos, Canary Islands, Spain, and at 
the 4.1 m Southern Astrophysical Research Telescope, Cerro Pach\'on, Chile.}
\altaffiltext{2}{Dipartimento di Astronomia, Universit\`a di Bologna, Bologna, 
Italy; mariaida.moretti@studio.unibo.it}
\altaffiltext{3}{INAF, Osservatorio Astronomico di Capodimonte, Napoli, Italy,
dallora@na.astro.it, ripepi@na.astro.it} 
\altaffiltext{4}{INAF, Osservatorio Astronomico di
Bologna, Bologna, Italy; gisella.clementini@oabo.inaf.it}
\altaffiltext{5}{INAF, Centro Galileo Galilei \& Telescopio Nazionale Galileo, S.
Cruz de La Palma, Spain; difabrizio@tng.iac.es}
\altaffiltext{6}{Department of Physics and Astronomy, Michigan State University, East Lansing, 
MI 48824-2320, USA; smith@pa.msu.edu, kuehncha@pa.msu.edu, beers@pa.msu.edu}
\altaffiltext{7}{Department of Astronomy, University of Florida, 211 Bryant Space Science Center, Gainesville, FL 32611-2055
; ndelee@astro.ufl.edu}
\altaffiltext{8}{Pontificia Universidad Cat$\rm{\acute{o}}$lica de Chile,
Departamento de Astronom\'{\i}a y Astrof\'{\i}sica, Santiago, Chile; mcatelan@astro.puc.cl}
\altaffiltext{9}{On sabbatical leave at Michigan State University, 
Department of Physics and Astronomy, 3215 Biomedical and Physical 
Sciences Bldg., East Lansing, MI 48824} 
\altaffiltext{10}{John Simon Guggenheim Memorial Foundation Fellow} 
\altaffiltext{11}{Joint Institute for Nuclear Astrophysics, Michigan State 
University, East Lansing, MI 48824, USA}
\altaffiltext{12}{Universidad de Concepci\'on, Departamento de
F\'{\i}sica, Concepci\'on, Chile, kkinemuchi@astro-udec.cl}


\begin{abstract} We present the first $V, B-V$ color$-$magnitude 
diagram of the Leo\,IV dwarf spheroidal galaxy, 
a faint Milky 
Way satellite recently discovered by the Sloan 
Digital Sky Survey. 
We have obtained $B,V$ time-series photometry reaching about half 
a magnitude below 
the Leo\,IV turnoff, which we detect at  $V$= 24.7 mag, and have 
performed the first study of the variable star population. We have identified three RR Lyrae stars (all fundamental$-$mode pulsators, RRab) 
and one SX Phoenicis variable in the galaxy.
In the period$-$amplitude diagram 
the Leo\,IV RR Lyrae stars are located close to the loci of 
Oosterhoff type I systems and the 
evolved fundamental-mode RR Lyrae stars in the Galactic globular cluster M3. However, 
their mean pulsation period, $\langle P{\rm ab}\rangle$=0.655 days, would  
suggest an Oosterhoff type II classification for this galaxy.
The RR Lyrae stars trace very well the galaxy's horizontal branch, 
setting its average magnitude at 
$\langle V_{\rm RR}\rangle= 21.48 \pm 0.03$ mag 
(standard deviation of the mean). 
This leads to a distance modulus of $\mu_{0}=20.94 \pm 0.07$ mag, corresponding 
to a distance of $154 \pm 5$ kpc, by adopting for the Leo\,IV dSph a reddening $E(B-V) = 0.04 \pm 0.01$ mag
and a metallicity of [Fe/H] = $-2.31$ $\pm$ 0.10.

\end{abstract}



\section[]{Introduction}
Dwarf Spheroidal (dSph) galaxies (\cite{Mat98}) 
provide important 
constraints on $\Lambda$-Cold-Dark-Matter ($\Lambda$CDM) theories  
of galaxy formation, which 
predict that several hundred 
small dark halo satellites should surround 
the halos of large galaxies like
the Milky Way (MW) and M31 
(\cite{Kly99}, \cite{Moo99}), and that dSphs
are the best candidates for the ``building blocks'' from which the MW and M31 were
assembled \citep{SZ78}.
Indeed, there is a sizeable discrepancy 
between $\Lambda$CDM theory and observations,
known as the ``missing satellite problem'' (\cite{Kau93}, \cite{Tol08}),
since the number of satellites predicted by theory is much higher than
the number of dSphs actually observed to surround the MW.
Solving the ``missing satellite problem'' would require the discovery of 
many new dSph satellites of our Galaxy (e.g., \cite{Wal09}).

The dSph galaxies surrounding the MW can be divided into two 
 groups:
``bright'' dSphs, mainly discovered before 2005, and ``faint'' dSphs, 
discovered in the last couple of years primarily from analysis of imaging obtained by the Sloan Digital Sky Survey 
(SDSS; \citealt{Yor00}). 
Bright and faint dSphs lie in two separate regions
in the absolute magnitude versus half-light radius plane
(see Figure~8 of Belokurov et al. 2007, hereafter B07). 

The bright dSphs include 10 systems (Draco, Ursa Minor, Fornax, Carina, Sculptor, Leo\,I, Leo\,II,
Sextans, Sagittarius, and Canis Major; 
\cite{Mat98}; \cite{Iba95}; 
\cite{Irw95}; \cite{Mar04}).  
These galaxies generally are found to contain stars exhibiting different chemical compositions than the stars in the Galactic halo
(see \cite{Hel06}, and references therein). 
Furthermore, they generally host RR Lyrae stars with pulsation 
properties that differ from the properties of the variables in the Galactic Globular Clusters (GCs), 
being ``Oosterhoff-intermediate'' (\cite{Oos39}; 
\cite{Cat09} and references therein). 
These two properties suggest that it is unlikely that the halo of
the MW was formed from objects with properties similar to the bright dSphs that
are observed today.
 
Since 2005, 15 new faint (effective surface brightnesses $\mu_{V}$$\gtrsim$ 28 arcsec$^{-2}$) dSph satellites of the MW 
have been discovered, primarily from SDSS imaging: 
Willman\,I, Ursa Major\,I, Ursa Major\,II,  
Bootes\,I, Coma Berenices (Coma), Segue\,I, Canes Venatici\,I (CVn\,I), Canes Venatici\,II (CVn\,II), Leo\,IV,  Hercules, 
Leo\,T, Bootes\,II, Leo\,V, Bootes\,III and Segue\,II (\citealt{Wil05a, Wil05b, Zuc06a, Gri06, Zuc06b, 
Bel06, Bel07, Irw07, Wal07,Bel08,Gri09, Bel09}). 
These faint galaxies 
have
 high  mass-to-light ratios  
and (often) distorted morphologies, probably due to tidal interactions with the MW.
They all host an ancient stellar population with 
chemical properties similar to that of external Galactic halo stars 
(\citealt{SG07,Kir08,Fre09}). 
Several of the faint dSphs have 
mean metallicities as low or lower than the most
metal-poor GCs and, generally, much lower than those 
of the bright dSphs.

Galactic GCs that contain significant numbers of RR
Lyrae stars have fundamental-mode (RRab) pulsators with mean
period ($\langle P{\rm ab}\rangle$) either around 0.55 days or around 0.65
days, and separate into the so-called Oosterhoff\,I (Oo\,I)
and Oosterhoff\,II (Oo\,II) types (\cite{Oos39}).
Extragalactic GCs and field RR Lyrae stars
in the bright dSph galaxies generally have, instead, 
$\langle P{\rm ab}\rangle$ intermediate between the 
two types 
(\cite{Cat09} and references therein). 
Four of the faint dSphs have been searched for variable stars so far 
(namely, Bootes\,I, \cite{Dal06}, \cite{Sie06}; CVn\,I, \cite{Kue08}; CVn\,II, \cite{Gre08};   
and Coma, \cite{Mus09}) and, with the exception of CVn\,I, were 
found 
to contain RR Lyrae stars with properties resembling those
of the MW Oo II  GCs.

All of the above characteristics suggest
that a much larger population of objects similar to the presently observed faint
dSphs may have been the ``building blocks" of the halos of large galaxies such
as the MW. The association is particularly clear with the outer halo of the MW,
which Carollo et al. (2007) have demonstrated to exhibit a peak metallicity of
[Fe/H] $= -2.2$, substantially lower than the inner halo (with a peak at [Fe/H]
$= -1.6$), and which is the dominant population at Galactocentric distances
beyond 15-20 kpc.

The Leo\,IV galaxy (R.A. = 11$^h$ 32$^m$ 57$^s$, 
decl. = $-00^{\circ}$ 32$^{\prime}$ 00$^{''}$, J2000.0; 
$l = 265^{\circ}.4$, $b = 56^{\circ}.5$) 
is one of the newly discovered SDSS dSphs, 
with absolute magnitude M$_V$ = $-5.1 \pm 0.6$ mag (B07)  
and surface brightness $\mu_{V}$ = 28.3 
mag arcsec$^{-2}$ \citep{SG07}. 
It is a low-mass [M = (1.4 $\pm$ 1.5) $\times$ $10^6$ \msol; \cite{SG07}] 
system, with half-light radius 
$r_h$ $\sim$ 3.3 arcmin (B07).
It is located at heliocentric distance
160$^{+15}_{-14}$ kpc with a position angle 
of 355$^{\circ}$ (B07).  
Its CMD is more complex than the CMDs of other galaxies 
discovered by Belokurov et al., due to the presence
of an apparently ``thick" red giant branch (RGB) and a blue horizontal branch (HB). 
The RGB thickness suggests the presence of stellar
populations of different age/metallicity (B07). 
There is no published CMD of
Leo\,IV other than the $i$, $g-i$ CMD that reaches $i \sim 22$ mag obtained from 
the SDSS discovery data, nor has a study of the variable stars in the galaxy yet been perfomed. 
\citet{SG07} obtained spectra for 18 bright stars in Leo\,IV
from which they derived an average velocity dispersion of 3.3 $\pm$ 1.7 km/s and
an average metallicity $\langle [Fe/H] \rangle = -2.31 \pm 0.10$ 
with a dispersion $\sigma{\rm [Fe/H]}$ = 0.15 dex, 
on the Zinn \& West (1984) metallicity scale (hereafter ZW84).
Kirby et al. (2008), using Keck DEIMOS spectroscopy coupled with spectral synthesis, 
measured the metallicity of a subset of 12 stars extracted from the Simon 
\& Geha (2007) sample. They obtained an average metallicity $\langle[Fe/H]\rangle = -2.58 \pm 0.08$,
 with a dispersion $\sigma{\rm [Fe/H]}$ = 0.75 dex and individual metallicities as low as [Fe/H] $\sim$ $-3.0$.

In this Letter we present 
the first $V, B-V$
CMD of the Leo IV dSph, reaching a depth of $V \sim$ 25.5 mag, sufficient to identify
the galaxy's main-sequence turnoff at $V\sim$ 24.7 mag.
We carry out a search for variable stars, and identify 
three fundamental-mode RR Lyrae stars 
(RRab) and one SX Phoenicis (SX Phe) variable.
We obtained $B$, $V$  light curves for each variable 
star and use the average magnitude of the  
RR Lyrae stars to estimate the distance to the galaxy.

\section{Observations and Data Reduction}
Time-series $B$, $V$, $I$ photometry of the Leo\,IV dSph
galaxy was collected on 2007, April 20-23, with the 
Wide Field Camera (WFC), the prime focus mosaic CCD 
camera of the 2.5 m Isaac Newton Telescope (INT), on 2007, 
May 11-12, with the Prime Focus Imaging Platform (PFPI) of the 
4.2 m William Hershel Telescope (WHT), and on 2007, March-May,
with the SOAR Optical Imager (SOI) of the 4.1 m 
SOuthern Astrophysical Research telescope (SOAR).
The fields of view (FOVs) covered by the three instruments 
are: 5.24 $\times$ 5.24 arcmin$^2$ for SOI at the SOAR telescope, 
16.2 $\times$ 16.2 arcmin$^2$ for PFPI at the WHT, 
and 33 $\times$  33 arcmin$^2$ for WFC at the INT.
We needed two partially overlapping SOI fields 
to cover the galaxy, while just one PFPI field was
sufficient, and from the INT data we could also infer
additional information on stars outside the Leo\,IV
half-light radius.
We obtained a total number of 37 $B$, 42 $V$ and 12 $I$ images 
of the galaxy. In this Letter we present results from the analysis
of the $B$ and $V$ data. 
Images  were reduced following standard procedures
(bias subtraction and flat$-$field correction) with 
IRAF. 
The INT and WHT data were corrected for linearity 
following recipies provided in the telescope's web pages.
We then performed PSF photometry using the 
DAOPHOT/ALLSTAR/ALLFRAME packages (\citealt{Ste87, Ste94}). 
Typical internal errors of the $B,V$ single-frame photometry for stars at
the HB level range from 0.02 to 0.03 mag for the INT and WHT data, and are of about 0.02 mag for the SOAR data.
The absolute photometric calibration was obtained 
using observations of standard stars in the Landolt (1992) fields
SA\,101, SA\,107, SA\,110 and PG1323, as extended by 
P.B. Stetson\footnote{See http://cadcwwwdao.nrc.ca/standards.},
which were observed at the INT during the night of 2007, April 22.
Errors of the absolute photometric calibration are
$\sigma_B=0.01$ mag, $\sigma_V=0.01$ mag, respectively.

\section{Identification of the Variable Stars}

\begin{figure} 
\includegraphics[width=16.3cm,bb=43 170 570 560,clip]{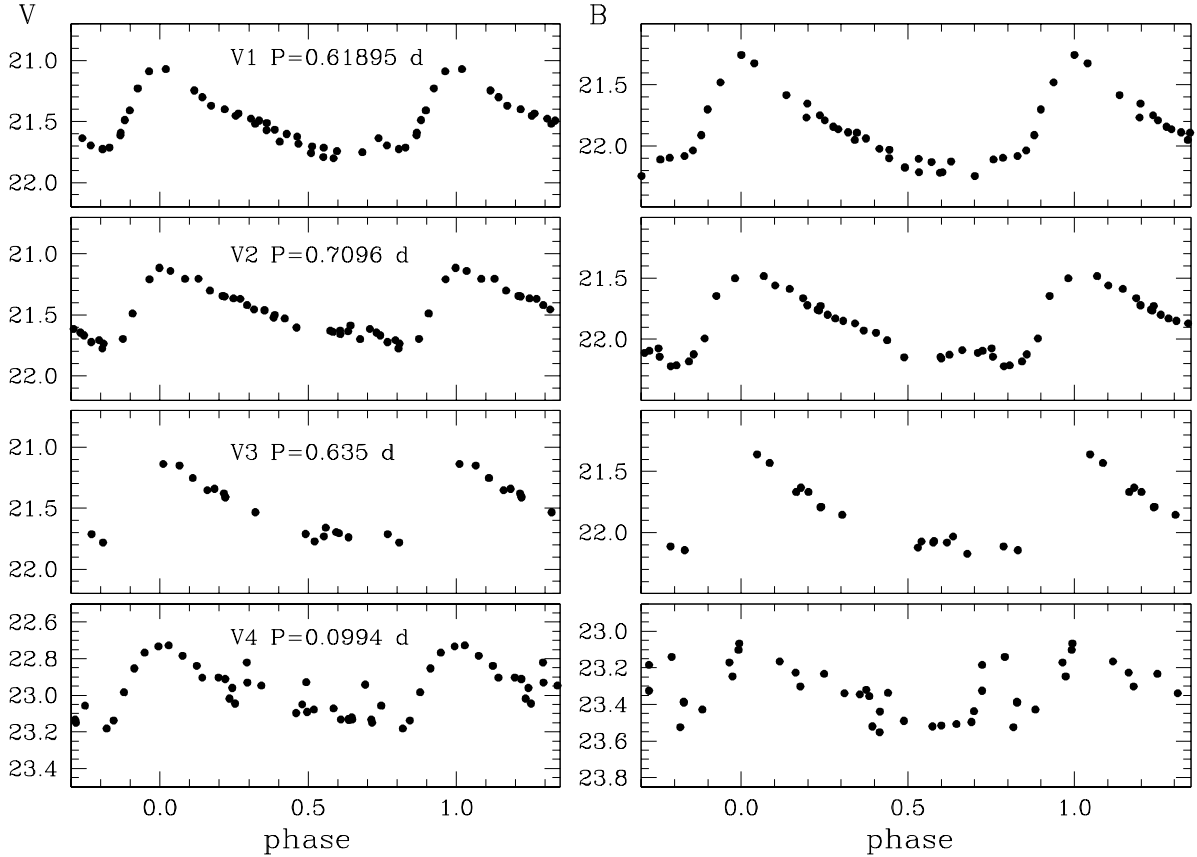}
\caption{$V$ (left panels) and $B$
(right panels) light curves of the variable stars discovered 
in the Leo\,IV dSph galaxy. {\it Three upper rows:} 
fundamental-mode RR Lyrae stars; {\it bottom row}: SX Phe variable.}
\label{fig:curve_di_luce}
\end{figure}

Variable stars were identified using the $V$ 
and $B$ time-series data separately.
First we calculated the Fourier transforms (in the Schwarzenberg-Czerny
1996 formulation) of the stars having at least 12 measurements
in each photometric band, then we averaged these transforms to 
estimate the noise and calculated the signal-to-noise
ratios (S/Ns).
Results from the $V$ and $B$ datasets were 
cross-correlated, and all stars with S/N$>$ 5
in both photometric bands were visually inspected, 
for a total of about 2000 objects.
We also checked whether some of the stars in the
Blue Straggler Stars (BSSs) region might be pulsating variables of 
SX Phe type.
Study of the
light curves and period derivation 
were carried out using the Graphical Analyzer of Time Series package (GRaTiS; \citealt{Cle00}). 
We confirmed the variability
and obtained reliable periods and light curves for 3
RR Lyrae stars, all fundamental-mode pulsators (stars: V1, V2 and V3), and for 
one SX Phe variable (star V4).
The identification and properties of the confirmed 
variable stars are summarized in Table 1, their light
curves are shown in Figure~\ref{fig:curve_di_luce}. 
The light curve data of the variable stars 
 and the photometric data of the galaxy CMD
are available on request
from the first author.

Stars V1, V2 and V4 lie inside the galaxy half-light radius,  
while V3 lies outside, at about 12 arcmin from the Leo\,IV center (see lower panel of Figure~4).
In the CMD, the SX Phe star is located in the region
of the BSSs, while all the RRab stars (V3 included) fall near the
galaxy's HB.
We checked the position of V4 on the period$-$luminosity
($PL$) relation of the SX Phe stars. 
Using the star's period and the absolute magnitude inferred from the  
apparent magnitude and the distance provided by the RR Lyrae stars 
(see Section~4), we found that V4 lies very close to the \citet{Por08}  $PL$ relation
for SX Phe stars, 
thus confirming the classification as an SX Phe star.

\begin{figure} 
\includegraphics[width=16.3cm,bb=20 146 578 708,clip]{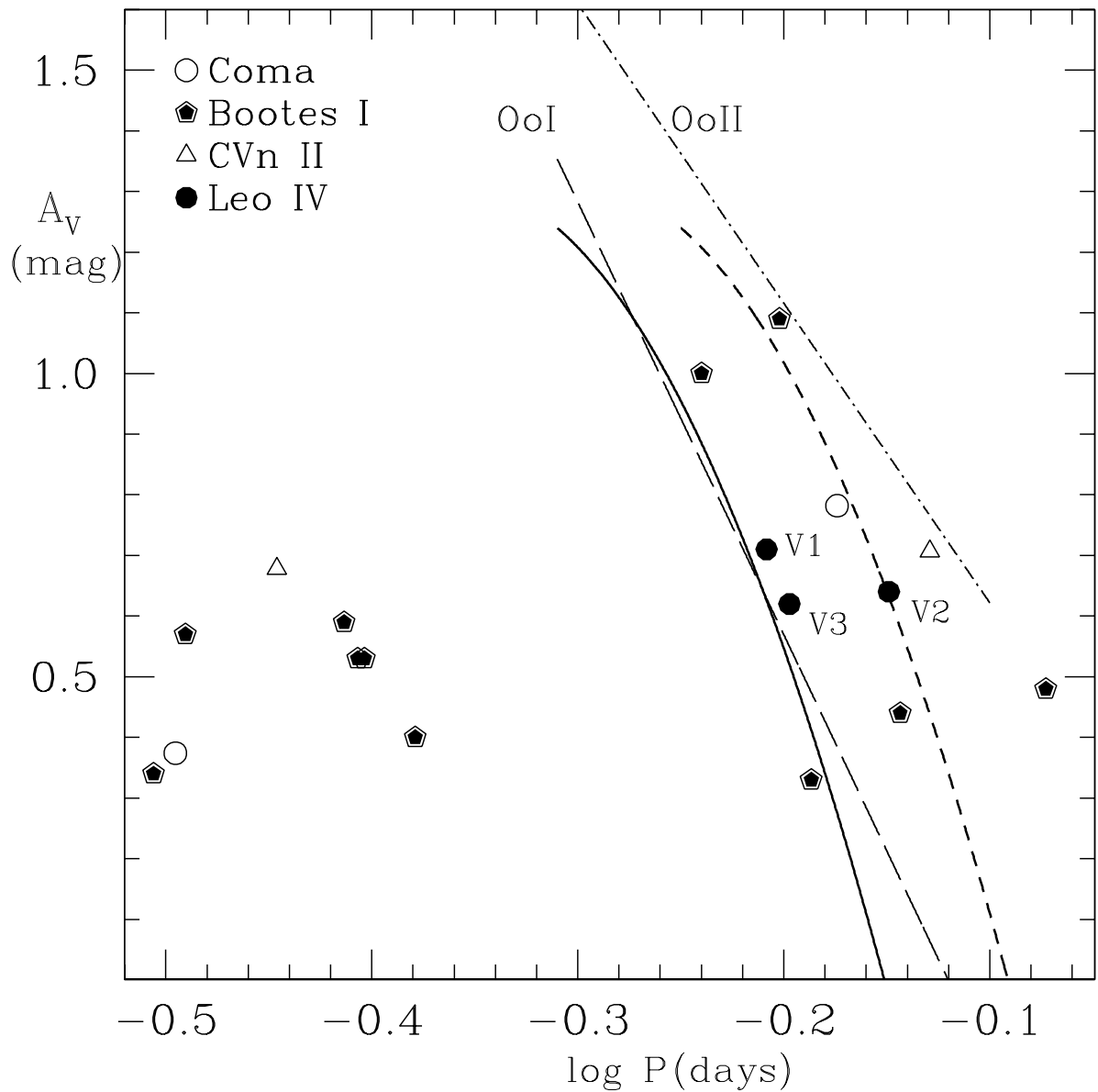}
\caption{$V$-band period-amplitude diagram of RR Lyrae stars in the 
Coma, Bootes\,I, CVn\,II, and Leo\,IV dSphs. Variables with 
$\log P > - 0.35 $ days are RRab pulsators, those with $\log P < - 0.35 $ days
are first-overtone (RRc) pulsators.
Long-dashed and 
dot-dashed lines show the position of the Oo\,I and Oo\,II Galactic
GCs, according to \citet{CR00}.
Period-amplitude distributions of the \emph{bona fide} regular 
(\emph{solid curve}) and well evolved (\emph{dashed curve}) 
RRab stars in M3, from Cacciari et al. (2005),
are also shown for comparison.}
\label{fig:baileynew}
\end{figure}

So far, only 4 dSphs of Oosterhoff\,II type are known, namely Ursa Minor among the bright
companions of the MW, and Bootes\,I 
(\cite{Dal06}; \cite{Sie06}), CVn\,II (\cite{Gre08}) and
Coma \citep{Mus09}, among the faint SDSS dSphs. 
The average pulsation period of the Leo\,IV RRab stars, $\langle P{\rm ab}\rangle$=0.655 days, would  
suggest that Leo\,IV too is more similar to the Oosterhoff type II systems. However, 
in the $V$-band period-amplitude diagram (see  Figure~\ref{fig:baileynew})
the Leo\,IV RRab stars fall close to the locus of  Oo\,I
systems (from \cite{CR00}), with V1 and V3 lying near 
 the distribution 
of the \emph{bona fide} regular variables, and V2 lying close to the locus of the
 well evolved fundamental-mode RR Lyrae stars in the Galactic GC M3 (from Cacciari, Corwin \& Carney 2005).
Nevertheless, V2 does not appear to be overluminous in the CMD, as would be required if the star were evolved off the
zero-age HB. 
The ambiguous behavior and the small number of variable stars make the conclusive assignment
of an Oosterhoff type to the Leo\,IV dSph rather difficult.

We used the parameters of the 
Fourier decomposition of the $V$-band
light curve, along with the \citet{JK96} 
method for RRab stars, to estimate the metallicity on the ZW84 scale of V1, the only variable of our RR Lyrae 
sample which satisfies the \citet{JK96} regularity conditions.
The metallicity we derived for V1 
is in good agreement
with the spectroscopic metallicity derived for another RR Lyrae star
 (variable V2) by \citet{Kir08} (see last column of Table~1).

\section{The CMD, Structure, and Distance of Leo IV} 

Figure~3 shows the $V$, $B - V$ CMD
of the Leo\,IV dSph obtained by plotting all stellar-like objects located within the 
half-light radius of 3.3 arcmin from the B07 center of Leo\,IV.  
The selection between stars and galaxies,
for magnitudes brighter than  $V$=22.5 mag, was done with the software
 Source Extractor (SExtractor;  \cite{BA96}). 
Variable stars are plotted in the CMD according
to their intensity-averaged magnitudes and 
colors (see Table~1), using filled circles for stars V1, V2 and V4, and a cross for V3. 
Although well outside the Leo\,IV half-light radius, V3  
appears to be perfectly located on the galaxy's HB, thus
confirming its membership to the galaxy.
%
Non-variable HB stars 
are marked by (blue) filled circles.
The CMD reaches $V$ $\sim$ 25.5 mag, and appears to be heavily contaminated at
every magnitude level by field objects belonging to the MW. 
We used the mean ridgeline of the Galactic GC M15 
(from \cite{DH93}; solid line)
properly shifted in magnitude and color, and selected 
as stars most likely belonging to the Leo\,IV galaxy the sources lying
within $\pm$0.05 mag from the ridgeline of M15 (red dots). 
To allow for the larger photometric errors, 
for magnitudes fainter than $V$=23.5 mag, we also considered as belonging to the galaxy
stars with $B-V$ color in the range from 
  $\pm$ 0.05 mag and $\pm$ 0.10 mag 
from the ridgeline of M15 (cyan dots). 
The HB of Leo\,IV shows up quite clearly and,  
along with the galaxy's RGB, is well reproduced by the 
ridgeline of M15, implying that
Leo\,IV has an old and metal-poor
stellar population with metallicity comparable to
that of M15 ([Fe/H] = $-2.15$ $\pm$ 0.08, on the ZW84  scale). 
We also note that, by adopting for M15 a reddening value of 
$E$($B - V$)= 0.10 $\pm$ 0.01 mag \citep{DH93}, 
the color shift needed 
to match the HB and RGB of Leo\,IV implies for the galaxy a reddening of 
$E$($B - V$)= 0.04 $\pm$ 0.01 mag. For comparison, the reddening in the direction of Leo\,IV 
obtained from the \citet{Sch98}  maps
is 0.025 $\pm$ 0.026 mag.
The objects marked by open circles are stars 
within the Leo\,IV half-light radius, whose membership to the galaxy was
confirmed spectroscopically by Simon \& Geha (2007). 
They include the RR Lyrae star V2 and 
a number of HB and RGB stars which fall very close to the M15 ridgeline, thus  
supporting our identification of the
Leo\,IV member stars.

The upper panel of Figure~4 shows a map of all sources observed in the FOV of the WHT observations
that we consider to belong to the Leo\,IV
galaxy, according to their position with respect to the M15 ridgeline, or with membership spectroscopically
confirmed by \citet{SG07} (open circles).
Symbols and color$-$coding are the same as in Figure~3 and, for 
non-variable stars, 
the symbol sizes are proportional
to the star's brightness. 
The solid circle shows the region corresponding to the 
half-light 
radius of Leo\,IV centered on the  
B07 coordinates for the galaxy.
The lower panel of Figure~4 shows a map of the sources observed in the INT 
FOV which lie 
within $\pm$ 0.05 mag in $B-V$ (for $V>$ 23.5 mag, within 
$\pm$ 0.10 mag in $B-V$) from the ridgeline of M15.
An overdensity of objects rather extended and irregular in shape is 
visible, corresponding to the region occupied by the Leo\,IV dSph. The black circle shows the 
 half-light radius of Leo\,IV, according to 
 B07. 
 The peripheral location of V3 is remarkable, and 
 provides further hints on the elongation and rather deformed morphology of the Leo\,IV dSph.
%

The average apparent magnitude of the galaxy's RR Lyrae stars 
is $\langle V_{\rm RR}\rangle = 21.48 \pm 0.03$ mag 
(standard deviation of the mean).
Assuming $M_V$ = 0.59 $\pm$ 0.03 mag for 
the absolute luminosity of the RR Lyrae stars at 
[Fe/H] = $-1.5$ \citep{cc03},
$\Delta$$M_V$/$\Delta$$[Fe/H]$ = 0.214 $\pm$ 0.047 mag dex$^{-1}$
for the slope of the luminosity-metallicity relation of 
RR Lyrae stars (\cite{Cle03}), 
$E(B-V)$ = 0.04 $\pm$ 0.01 mag and [Fe/H] = $-2.31$ (Simon \& Geha, 2007), the distance modulus of Leo\,IV
is $\mu_0$ = 20.94 $\pm$ 0.07 mag which corresponds to a distance 
$d$ = 154 $\pm$ 5 kpc. The error includes uncertainties in the 
photometry, reddening, metallicity, and RR Lyrae absolute magnitude, but does not take into account 
evolution off the zero-age HB which might contribute an additional 0.05 mag uncertainty,
bringing the total error budget to 0.09 mag.
This new, precise distance estimate agrees very well with the distance of 
160$^{+15}_{-14}$ kpc derived by B07.

\section[]{Summary and conclusions} 
We have identified and obtained $B, V$ light 
curves for three fundamental-mode RR Lyrae stars (V1, V2 and V3)  
and one SX Phe variable (V4) in the Leo\,IV dSph galaxy. 
In the period$-$amplitude diagram V1 and V3 fall close to the
loci of Oo\,I Galactic GCs and \emph{bona fide} regular variables in the Galactic GC M3, while V2 lies close to  
the loci of Oo\,II  and well evolved M3 RRab stars. However, their average period, 
$\langle P{\rm ab}\rangle$=0.655 days, would  
suggest an Oosterhoff\,II 
classification for the galaxy.  
From the average magnitude of the galaxy's 
RR Lyrae stars, the distance modulus of the 
Leo\,IV dSph is $\mu_0$ = 20.94
$\pm$ 0.07 mag ($d=154\pm5~\rm{kpc}$). 
One of the RR Lyrae stars (V3) lies at about 12 arcmin from the galaxy center.
Nevertheless, the mean magnitude places the star
exactly on the galaxy HB and close to the other two RRab stars, thus
suggesting a significant elongation of the Leo \,IV galaxy. 

\acknowledgments
We thank Evan Kirby and Joshua Simon 
for sending us identification and individual metallicities for member stars of the
Leo\,IV dSph galaxy. 
Financial support for this study was provided 
by PRIN INAF 2006 (P.I.: G. Clementini). 
H.A.S. thanks 
the Center for Cosmic Evolution 
and 
the U.S. NSF for support under grant AST0607249. 
M.C. is supported by Proyecto 
Basal PFB-06/2007, by FONDAP Centro de Astrof\'{i}sica 15010003,  
by Proyecto FONDECYT Regular \#1071002, and by a John Simon Guggenheim 
Memorial  Foundation Fellowship. T.C.B. acknowledges partial support from grants PHY 02-16783 and PHY 08-22648:
Physics Frontier Center/Joint Institute for Nuclear Astrophysics (JINA), awarded
by the US National Science Foundation.

\begin{table*}
\tiny
\caption[]{Identification and properties of variable stars in the Leo\,IV dSph 
galaxy}
\label{t:table1}
$$
\begin{array}{lcclllccccc}
\hline
\hline
\noalign{\smallskip}
{\rm Name}&{\rm \alpha}&{\rm \delta}&{\rm Type}&~~~~P&{\rm Epoch (max)}&\langle V\rangle&
\langle B\rangle& A_V~~     &A_B~~&{\rm [Fe/H]_{ZW84}}\\
~~        &{\rm (2000)}&{\rm (2000)}&          & ~{\rm (days)}& ~($-$2450000)   & {\rm (mag)}            
& {\rm (mag)}  &{\rm (mag)}&{\rm (mag)}&\tablenotemark{a}\\
\noalign{\smallskip}
\hline
\noalign{\smallskip}
	    
{\rm V1}  & 11~32~59.2 & -00~34~03.6 & {\rm RRab}  & 0.61895 & ~~4212.453  & 21.47 & 21.82 & 0.73  & 0.99 &  -2.11 \\ 
{\rm V2}  & 11~32~55.8 & -00~33~29.4 & {\rm RRab}  & 0.7096  & ~~4214.543  & 21.46 & 21.86 & 0.64  & 0.76 &  -2.03 \\       
{\rm V3}  & 11~33~36.6 & -00~38~43.3 & {\rm RRab}  & 0.635   & ~~4212.453  & 21.52 & 21.81 & 0.65  & 0.82 &$\nodata$\\
{\rm V4}  & 11~32~45.4 & -00~31~44.4 & {\rm SX~Phe}& 0.0994  & ~~4213.397  & 22.96 & 23.34 & 0.37  & 0.38 &$\nodata$\\

\hline
\end{array}
$$
\tablenotetext{a}{The metallicity of V1 was derived from the Fourier parameters of the $V$-band 
light curve. The metallicity of V2 is from \citet{Kir08}.
}\\
\end{table*}

\begin{figure*}[h]
\includegraphics[width=16.3cm, height=16.3cm]{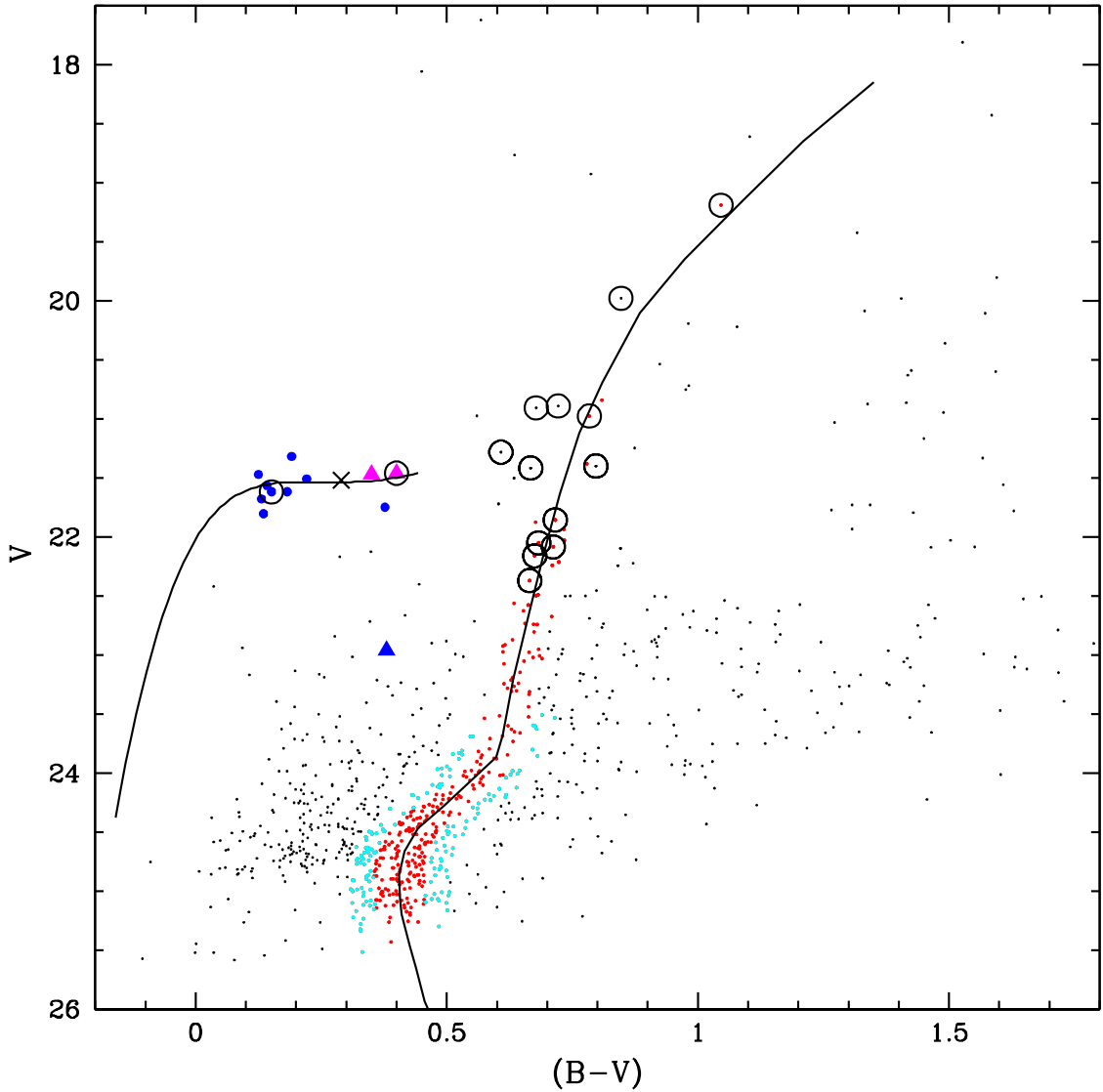}
\caption{$V$, $B - V$ CMD
of the Leo\,IV dSph obtained by plotting stellar-like objects located within 
the 
half-light radius of 3.3 arcmin. 
Variable stars V1, V2 and V4 are marked by triangles, star V3 by a cross, 
and non-variable HB stars by (blue) filled
circles. 
The solid line
is the ridge line of the Galactic GC M15.
Red and cyan dots are stars respectively within $\pm$ 0.05 mag in $B-V$ and, for $V>$ 23.5 mag,
 from $\pm$ 0.05 and $\pm$ 0.10 mag in $B-V$ from the ridgeline of M15. 
Open circles mark member stars of the Leo\,IV dSph according to Simon \& Geha (2007) and Kirby et al. (2008).
}
\end{figure*} 

\begin{figure}[h]
\includegraphics[width=9cm, height=9cm]{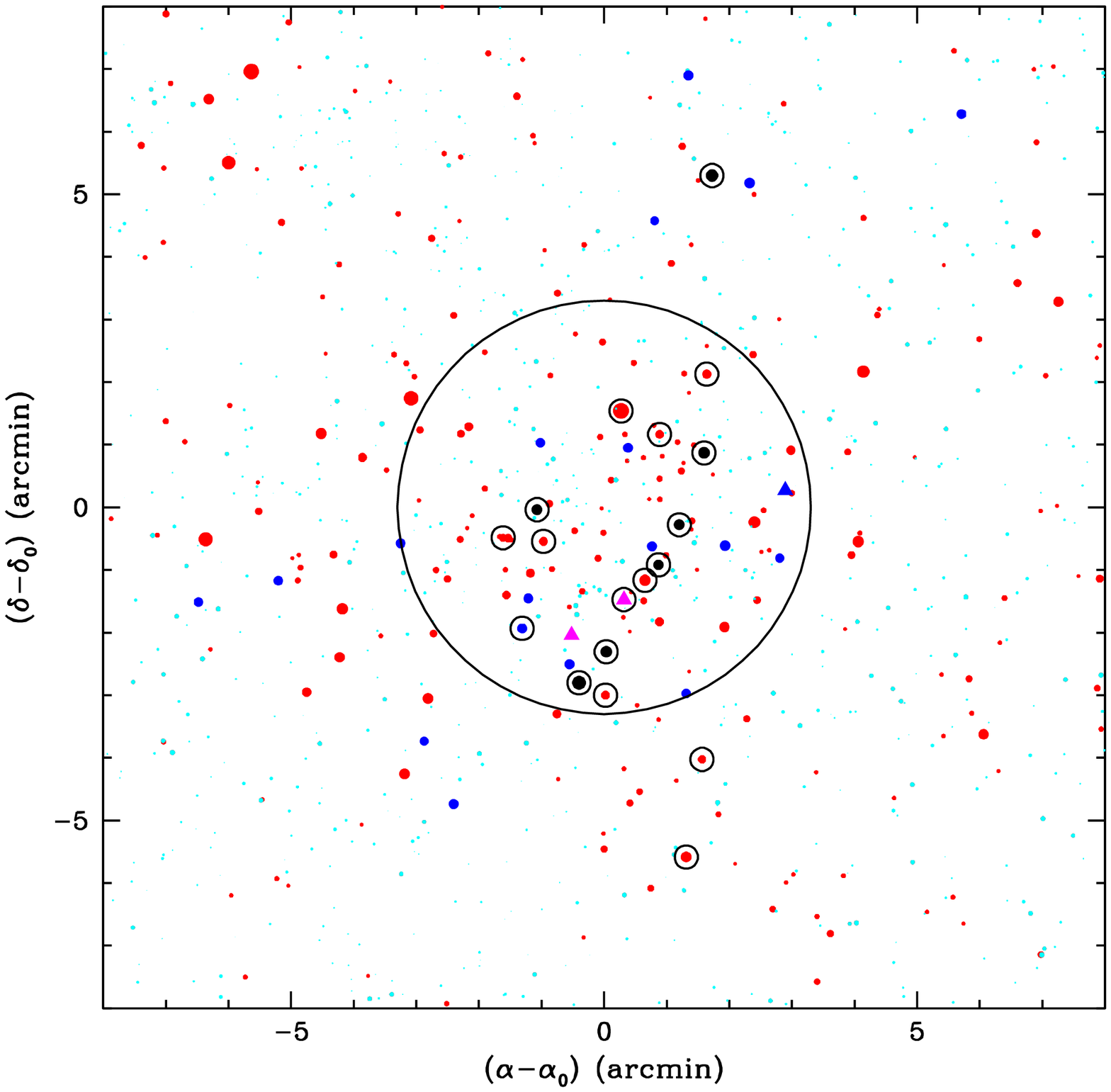}
\includegraphics[width=9cm, height=9cm]{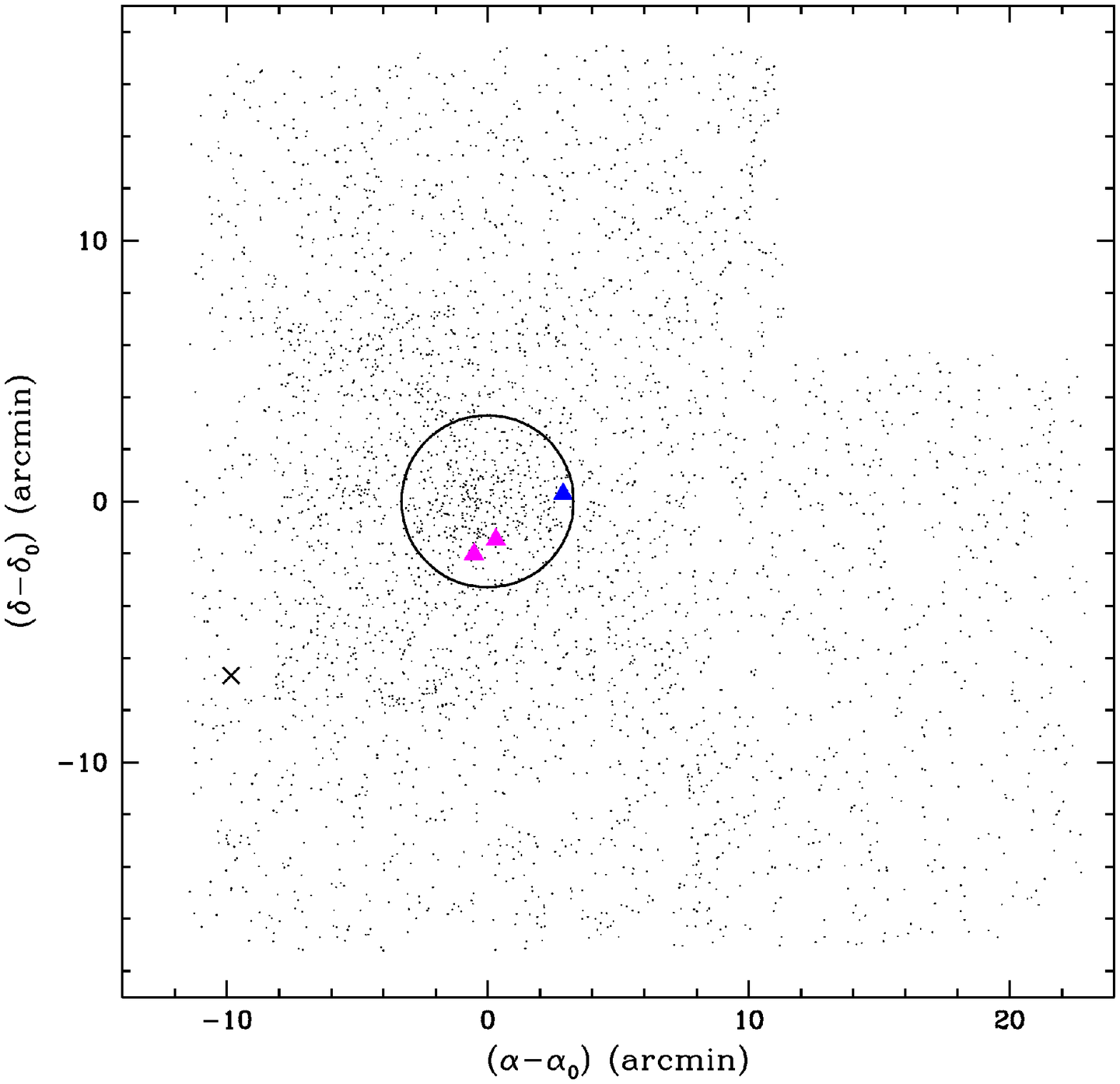}
\centering
\caption{{\it Upper panel:} Map of sources in the FOV of the WHT observations, which we 
consider to belong to the Leo\,IV galaxy according to the fit
with the M15 ridgeline, or
with membership spectroscopically
confirmed by \citet{SG07} (open circles).
Symbols and color-coding are the same as in Figure 3 and, for non-variable
stars, the symbol sizes are proportional to the star's brightness. 
{\it Lower panel:} Map of sources observed in the INT FOV which lie within $\pm$ 0.05 mag in $B-V$ (for $V >$ 23.5 mag, 
 within $\pm$ 0.10 mag in $B-V$)  from the ridgeline of M15.
The symbols and color$-$coding for the variable stars are the same as in Figure~3. 
In both panels the large circle shows the region corresponding to the 
half-light radius of Leo\,IV centered on the  
B07 coordinates for the galaxy.
}
\end{figure}

\end{document}